\documentclass[twocolumn, aps, prl, superscriptaddress]{revtex4-1}

\usepackage{graphicx,soul}
\usepackage{amsmath,amssymb,mathtools}
\usepackage{color}
\usepackage[colorlinks=true,linkcolor=blue,citecolor=blue]{hyperref}

\definecolor{highlightcolor}{rgb}{0.7,0.1,0.9}

\begin{document}

\title{\textcolor{black}{Persistent coherent beating in coupled parametric oscillators}}

\author{Leon Bello}
\affiliation{Department of Physics and BINA Center of Nanotechnology, Bar-Ilan University, 52900 Ramat-Gan, Israel}

\author{Marcello Calvanese Strinati}
\affiliation{Department of Physics, Bar-Ilan University, 52900 Ramat-Gan, Israel}

\author{Emanuele G. Dalla Torre}
\affiliation{Department of Physics, Bar-Ilan University, 52900 Ramat-Gan, Israel}

\author{Avi Pe'er}
\affiliation{Department of Physics and BINA Center of Nanotechnology, Bar-Ilan University, 52900 Ramat-Gan, Israel}

\date{\today}

\begin{abstract}
Coupled parametric oscillators were recently employed as simulators of artificial Ising networks, with the potential to solve computationally hard minimization problems. We demonstrate a new dynamical regime within the simplest network - two coupled parametric oscillators, where the oscillators never reach a steady state, but show persistent, full-scale, coherent beats, whose frequency reflects the coupling properties and strength. We present a detailed theoretical and experimental study and show that this new dynamical regime appears over a wide range of parameters near the oscillation threshold and depends on the nature of the coupling (dissipative or energy preserving). Thus, a system of coupled parametric oscillators transcends the Ising description and manifests unique coherent dynamics, which may have important implications for coherent computation machines.
\end{abstract}


\maketitle

In modern physics, the optical parametric oscillator (OPO) is widely known due to its applications in classical and quantum optics. Below the oscillation threshold, the OPO generates squeezed vacuum~\cite{PhysRevA.29.408,PhysRevA.30.1386,JOSAB.4.001465, Lvovsky_2015}, with applications in metrology~\cite{PhysRevD.23.1693,Harry2010,Aasi2013,qdm}, micro- and nano-electromechanical systems~\cite{PhysRevB.67.134302,PhysRevE.79.026203,PhysRevE.80.046202,PhysRevE.84.016212}, quantum information~\cite{Ralph1999,Furusawa1998,Braunstein2000,s41377-018-0011-3} and quantum communications~\cite{PhysRevA.61.010303,s41467-018-03083-5}. Above threshold, an OPO is the primary source of coherent light at wavelengths that are not laser accessible.

The working mechanism of a degenerate parametric oscillator is the well known period doubling instability~\cite{strogatz2007nonlinear}. In contrast to the lasing instability, the gain in a parametric oscillator depends on the phase of the oscillation, relying on the coherent nonlinear coupling between the pump field (at frequency \textcolor{black}{$\gamma$}) and the oscillation (at exactly \textcolor{black}{$\gamma/2$}) to amplify a single quadrature component of the oscillation field while attenuating the other quadrature. The phase of the amplified quadrature can acquire two distinct values, which give rise to two \textcolor{black}{inequivalent} solutions \textcolor{black}{with a relative shift of $\pi$}. Each solution breaks the time-translational symmetry of the pump, and thus an OPO is the simplest example of a classical discrete (Floquet) time crystal~\cite{PhysRevLett.116.250401,PhysRevLett.117.090402,PhysRevB.94.085112,PhysRevLett.118.030401,choi2017observation,sacha2017time,yao2018classical,sullivan2018dissipative,doi:10.1063/PT.3.4020,PhysRevLett.122.015701}. 

Borrowing the common terminology from condensed matter, a \textcolor{black}{single} parametric oscillator can be viewed as a classical two-level system (spin-1/2, or Ising spin). Based on this analogy, it has been recently suggested that coupled parametric oscillators can be used to simulate chains or networks of Ising spins~\cite{PhysRevA.88.063853,nphoton.2016.68,s41534-017-0048-9,takesue2018large2dising,king2018emulating,Hamerly:18,1805.05217,CerveraLierta2018exactisingmodel,Tiunov:19,arXiv:1903.07163,PhysRevLett.122.213902}. 
The Ising simulation relies on the inherent mode competition and positive feedback within the oscillators to find the most efficient (coupled-mode) oscillation, which can reflect the ground-state configuration of the corresponding Ising model (under certain assumptions): A set of optical parametric oscillators can therefore represent a set of independent spin-1/2 systems, where coupling of the optical field between the oscillators gives rise to a coupled network of spins. \textcolor{black}{Depending on the coupling, any two oscillators will prefer to phase-lock either in-phase (``ferromagnetic", $00$ or $\pi\pi$) or anti-phase (``anti-ferromagnetic", $0\pi$ or $\pi0$).}
This simulator, called coherent Ising machine (CIM), can simulate the spin dynamics and aims at calculating the ground state of the corresponding Ising model, thereby solving minimization problems that cannot be solved on a classical computer. 

\textcolor{black}{Here, we show that the dynamics of coupled parametric oscillators extends well beyond that of coupled Ising spins, and demonstrate a new dynamical regime of persistent coherent beating between the oscillators, that exists within a broad range of parameters near the oscillation threshold. We consider the simplest case of {\it two} coupled degenerate parametric oscillators with coupling that incorporates both energy-dissipating and energy-preserving components}, \textcolor{black}{and show that the latter induces a unique coherent dynamics, where the oscillators never \textcolor{black}{phase-lock}, but rather display everlasting, full-scale beats, \textcolor{black}{with a stable phase difference of $\pi/2$}. This is in contrast to usual wave phenomena, where coherent beats are normally a \textit{transient phenomenon} that decays due to decoherence, dissipation and non-linear effects.} Instead, \textcolor{black}{phase locking is induced by} the dissipative coupling component. The transition from the beating to the \textcolor{black}{phase-locked} regime by slow variation of the coupling properties may be of interest to coherent computing schemes.

We realize experimentally a pair of coupled parametric oscillators using parametrically driven radio-frequency (RF) resonators with a tunable coupling. Our experimental findings agree with the solution of an analytical model that accounts for periodic drive, gain, losses, nonlinearities, and coupling with energy-preserving and dissipative components. \textcolor{black}{The dissipative component simply reflects the possible imbalance of the couplings between the oscillators, where the coupling rate from oscillator A to B may be different from the coupling rate from B to A (indicating dissipation in the coupling channels)~\cite{PhysRevA.100.023835}.}

Our main finding is that, depending on the relation between the two coupling components, two distinct oscillation regimes exist: (i) When the dissipative component of the coupling dominates, the system \textcolor{black}{prefers phase-locking either in-phase or anti-phase}, which is the working principle of CIMs~\cite{PhysRevA.88.063853}; (ii) When the energy-preserving coupling dominates, the system displays a richer phenomenology: When the pump frequency is twice the bare-oscillator frequency, the system exhibits periodic beats that never decay or lose coherence. Only when the pump power is raised further, beyond a higher nonlinear threshold, the oscillators \textcolor{black}{phase-lock}. The beating regime, which is unique to parametric oscillators and cannot be observed in coupled lasers, represents a trajectory in phase space that visits periodically all the possible spin configurations and may have implications for the operation of CIMs. This novel regime, in which the system is not amenable to the description of Ising spins, is the main subject of our analysis.

\begin{figure}[t]
\centering
\includegraphics[width=8.6cm]{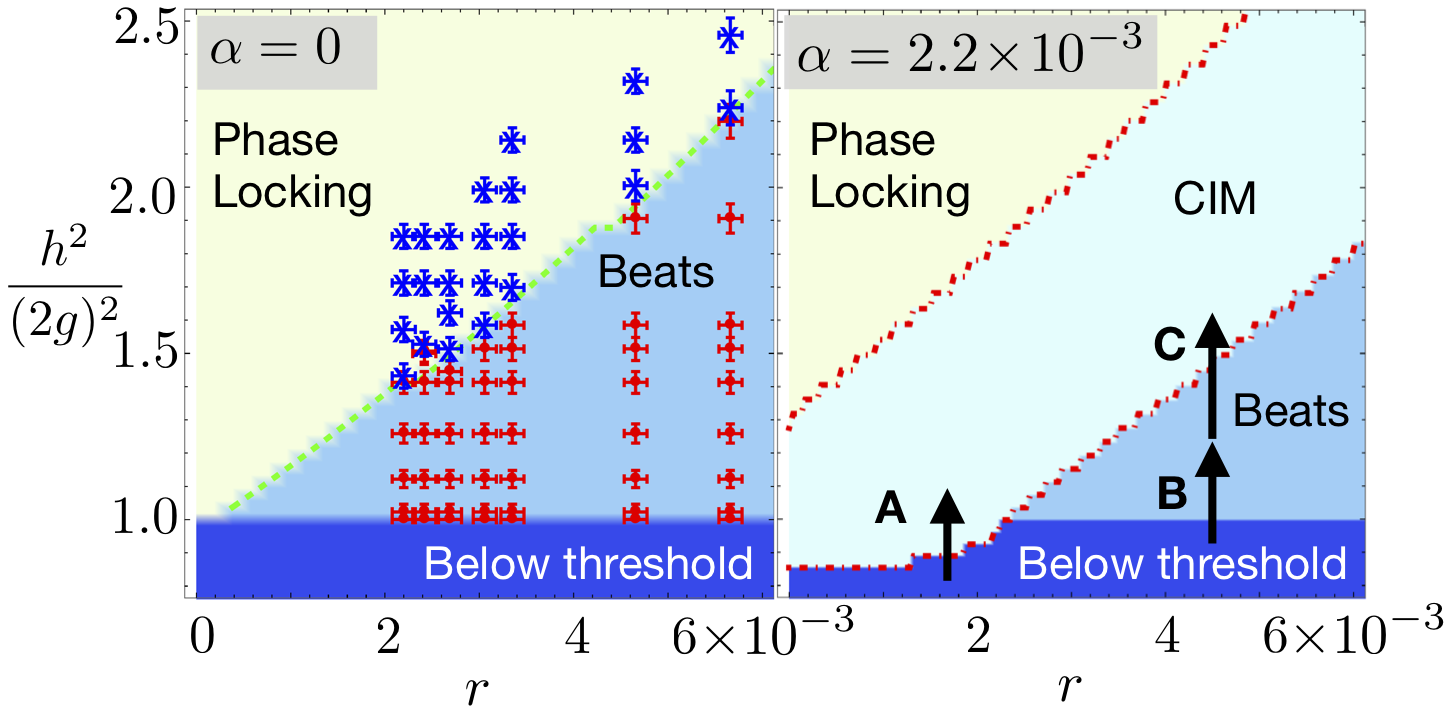}
\caption{Stability phase diagram in the $h^2/{(2g)}^2$ vs. $r$ plane, computed from Eq.~\eqref{eq:nonlinearmathieuequationflowcoupled} with \textcolor{black}{$g\!=\!3.2\!\times\!10^{-2}$} \textcolor{black}{and $\beta\!=\!10^{-2}$}. Left: {energy-preserving coupling only} ($\alpha\!=\!0$) and Right: With an {energy-dissipating coupling} of $\textcolor{black}{\alpha\!=\!2.2\!\times\!10^{-3}}$. Different colors indicate different phases \textcolor{black}{(see labels)}. For $\alpha\!=\!0$, the experimental points (red dots and blue crosses, indicating that beats or \textcolor{black}{phase locking} was experimentally observed, respectively) are superimposed on the theoretical phase diagram.}
\label{fig:floquetphasediagram}
\end{figure}

Theoretically, we first study the coupled system by resorting to a linear stability analysis, based on Floquet's theorem~\cite{magnus1979hill,chicone2006ordinary,1367-2630-17-9-093039}, which allows us to characterize all the parametric instabilities of the system without nonlinearities. We then employ a multi-scale expansion~\cite{kevorkian1996multiple}, \textcolor{black}{also known as slow-varying envelope approximation in non-linear optics}, to determine analytically the phase diagram of the coupled OPOs including nonlinearities. We find {four} major phases of oscillation (see Fig.~\ref{fig:floquetphasediagram}): (i) A stable phase of no oscillation below-threshold (semiclassical squeezed noise). (ii) A \textcolor{black}{CIM region} slightly above threshold with \emph{two} possible \textcolor{black}{phase-locked} oscillations. This CIM \textcolor{black}{region} exists only when the coupling is dominated by the dissipative component. (iii) Further above threshold, a \textcolor{black}{region} with \emph{four} possibilities of \textcolor{black}{phase-locked} oscillation. (iv) An extended region near threshold, where the oscillators show periodic exchange of energy between them (coherent beating) with a non-universal envelope (beat) frequency. This beating behavior, which appears only when the energy-preserving component of the coupling dominates, was not addressed before, and differs from the usual \textcolor{black}{description} of parametric oscillators, whose frequency is dictated by the pump only. The existence of the beating \textcolor{black}{region} near threshold suggests an alternative route to the \textcolor{black}{CIM behaviour}: In addition to the standard direct transition from sub-threshold to the \textcolor{black}{CIM region} (arrow \textbf{A} on Fig.~\ref{fig:floquetphasediagram}, right panel), the oscillators may also cross first into the beating \textcolor{black}{region} \textcolor{black}{(arrow \textbf{B})} and only then reach the \textcolor{black}{CIM phase-locked regime} (arrow \textcolor{black}{\textbf{C}}).

\textit{Theoretical model}. We study a system of two degenerate single-mode parametric oscillators, with equal gain and loss terms, coupled via energy-preserving and energy-dissipating terms, in the presence of pump-depletion nonlinearity. We analytically model our system by a set of classical equations of motion:
\begin{equation}
\begin{array}{l}
\ddot x_1+\Omega^2_1(t,0)\,x_1+\omega_0g\,\dot x_1-\omega_0{(r-\alpha)}\,\dot x_2=0\\
\ddot x_2+\Omega^2_2(t,\phi)\,x_2+\omega_0g\,\dot x_2+\omega_0{(r+\alpha)}\,\dot x_1=0
\end{array} \,\, .
\label{eq:equationsofmotions}
\end{equation}
Here, $x_1$ and $x_2$ represent the oscillation amplitudes,
the resonant frequency $\Omega_{1,2}(t)$ is parametrically modulated in time as
$\Omega^2_j(t,\phi)\!=\!\omega^2_0[1\!+\!h(x_j)\sin(\gamma t+\phi)]$ ($j=1,2$), with $\omega_0$ being the resonant frequency of the oscillators, $\gamma$ the pump frequency and $\phi$ the relative phase between the pumps; $h(x)\!=\!h(1\!-\!\beta x^2)$ represent the normalized pump power, where $\beta$ accounts for the pump depletion nonlinearity when the oscillation is substantial; $g$ is the intrinsic loss and {$r$ and $\alpha$ represent the energy-preserving and energy-dissipating coupling terms, respectively}.

If $x_{1,2}$ are sufficiently small, the nonlinearity can be neglected ($\beta\!=\!0$), which is valid \textcolor{black}{close to} the oscillation threshold, allowing us to diagonalize Eq.~\eqref{eq:equationsofmotions} by introducing the two eigenmodes {$x_\pm(t)\!=\!x_1(t)\!+\!\textcolor{black}{q}_\pm(r,\alpha)\,x_2(t)$, where the coefficients $\textcolor{black}{q}_\pm(r,\alpha)$ are determined by the values of $r$ and $\alpha$}. The stability analysis of the system can then be carried out by means of a perturbative approach based on Floquet's theorem. We {discuss} here the main results (for details, see~\cite{PhysRevA.100.023835}).

When the dissipative coupling dominates, $\alpha\!>~\!r$, there is only one parametric resonance at $\gamma\!=\!2\omega_0$. The two eigenmodes $x_\pm$ have different thresholds $h_{\rm th,\pm}\!\sim\! 2g\pm2\sqrt{\alpha^2\!-\!r^2}$. Therefore, by increasing $h$ above the lower threshold, one can selectively excite $x_-$, and for higher $h$, also $x_+$. The two modes are excited \emph{independently} and oscillate with the same frequency ($\gamma/2$), with an exponential time dependence: $x_\pm(t)\!\sim\! e^{(h-h_{\rm th,\pm})\omega_0t/4}\,\cos(\gamma t/2)$. This is the standard case for CIMs.

In contrast, when the energy-preserving coupling dominates, $r\!>\!\alpha$, the system displays a richer \textcolor{black}{phenomenology}: The coupling lifts the degeneracy between the oscillators and generates \textit{two} coupled modes $x_\pm$ with linear eigenfrequencies $\omega_\pm\!=\!\omega_0\left(1\!\pm\!\sqrt{r^2\!-\!\alpha^2}/2\right)$. Yet, parametric resonances of the coupled system appear in \textit{three} distinct frequencies: two resonances expectedly at $\gamma\!=\!2\omega_\pm$, which represent independent excitation of each coupled mode; and one new, less expected resonance at $\gamma\!=\!\omega_{+}\!+\!\omega_{-}\!=\!2\omega_0$, where both eigenmodes $x_\pm$ are excited \emph{simultaneously}, leading to full scale beats above the threshold $h_{\rm th}\!\sim\!2g$: $x_\pm(t)\!\sim\! e^{\mp i\omega_0t{\sqrt{r^2-\alpha^2}}/2}\,e^{(h-h_{\rm th})\omega_0t/4}\,\cos(\gamma t/2)$. Indeed, when pumped at $\gamma\!=\!2\omega_0$, the system cannot oscillate on a single coupled mode (due to the frequency mismatch), but non-degenerate oscillation of both modes is still possible. We therefore see (Fig.~\ref{fig:floquetphasediagram}) that at $r\!=\!\alpha$ the system undergoes a transition from a CIM to a coherent beating behaviour. The actual existence of the three resonances depends on the pump phase $\phi$: when $\phi\!=\!0$, only the resonance at $\gamma\!=\!2\omega_0$ can be excited, whereas for $\phi\!=\!\pi$ only the resonances at $\gamma\!=\!2\omega_\pm$ exist (for a generic $0\!<\!\phi\!<\!\pi$, all three resonances are found). 

We now expand the analysis further above the threshold (beyond the linear Floquet analysis) by incorporating the nonlinearity $\beta\!\neq\!0$ and resorting to a multiple-scale perturbative expansion~\cite{kevorkian1996multiple}. For brevity, we focus on degenerate pumping at $\gamma\!=\!2\omega_0$, where the system displays richer physics, and $\phi\!=\!0$.

\begin{figure}[t]
\centering
\includegraphics[width=8.2cm]{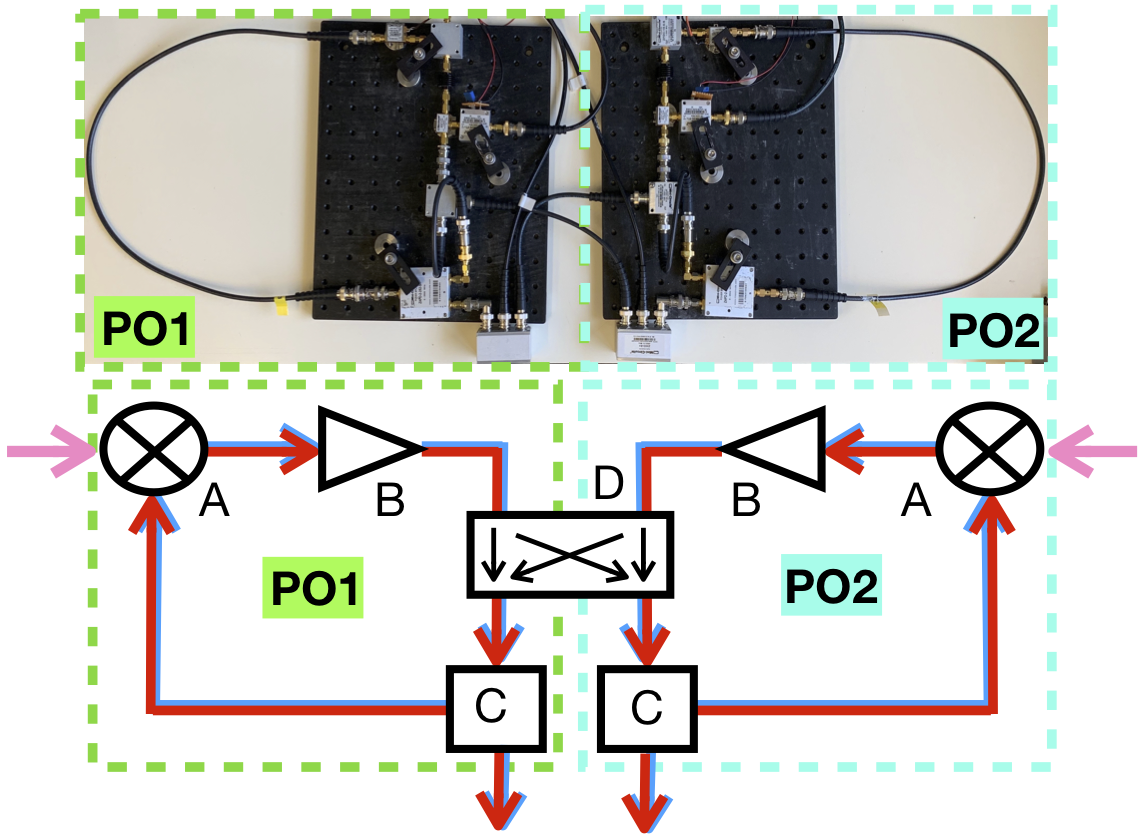}
\caption{(Top) Picture and (Bottom) scheme of the experimental setup. Our parametric oscillators are implemented in RF using standard components: (A) frequency mixer, (B) broadband amplifier, (C) coupler, (D) power splitter coupler.}
\label{fig:schemeoftheexperimentalapparatus}
\end{figure}

\begin{figure}[h!]
\centering
\includegraphics[width=8cm]{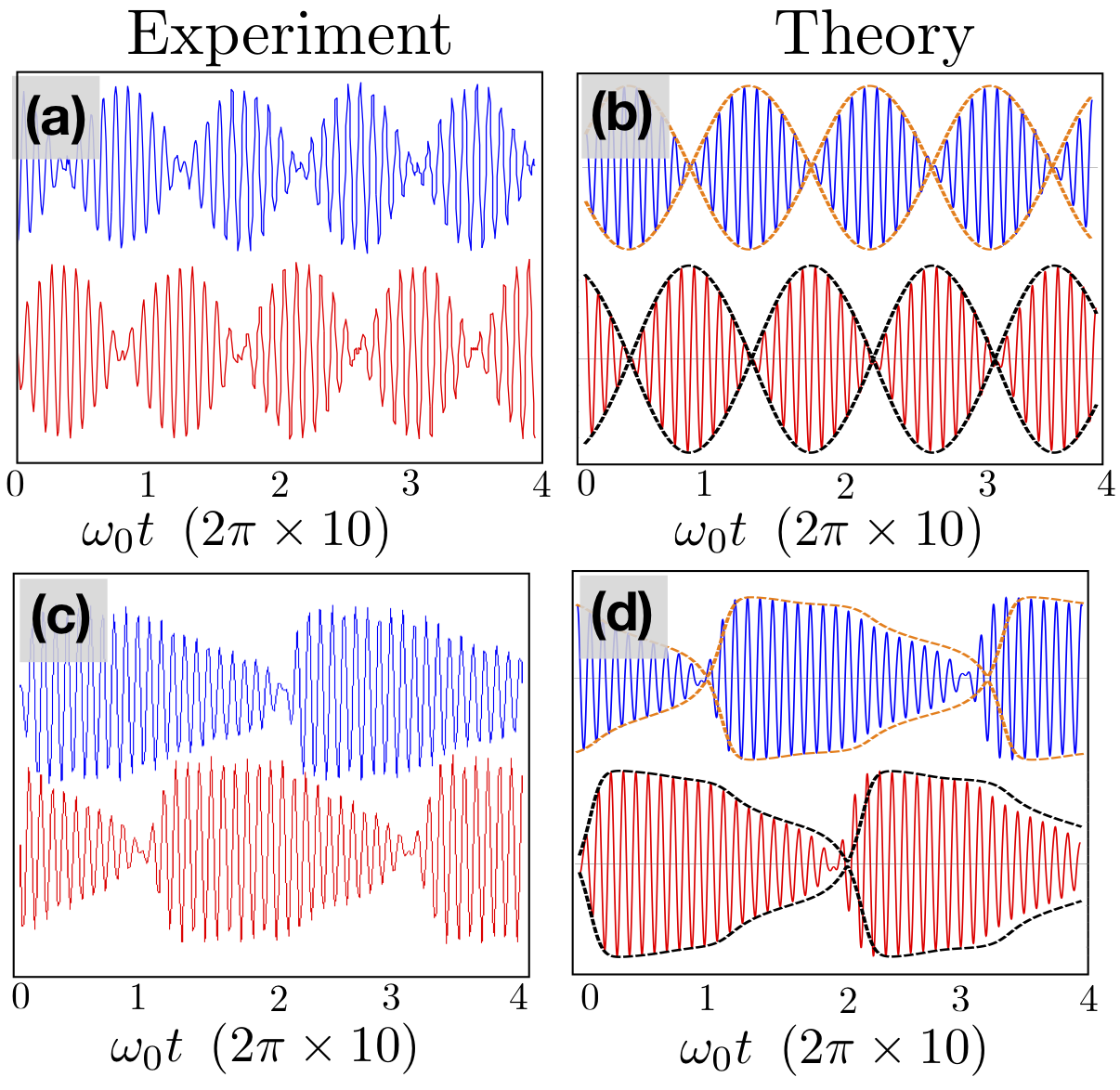}
\includegraphics[width=8.5cm]{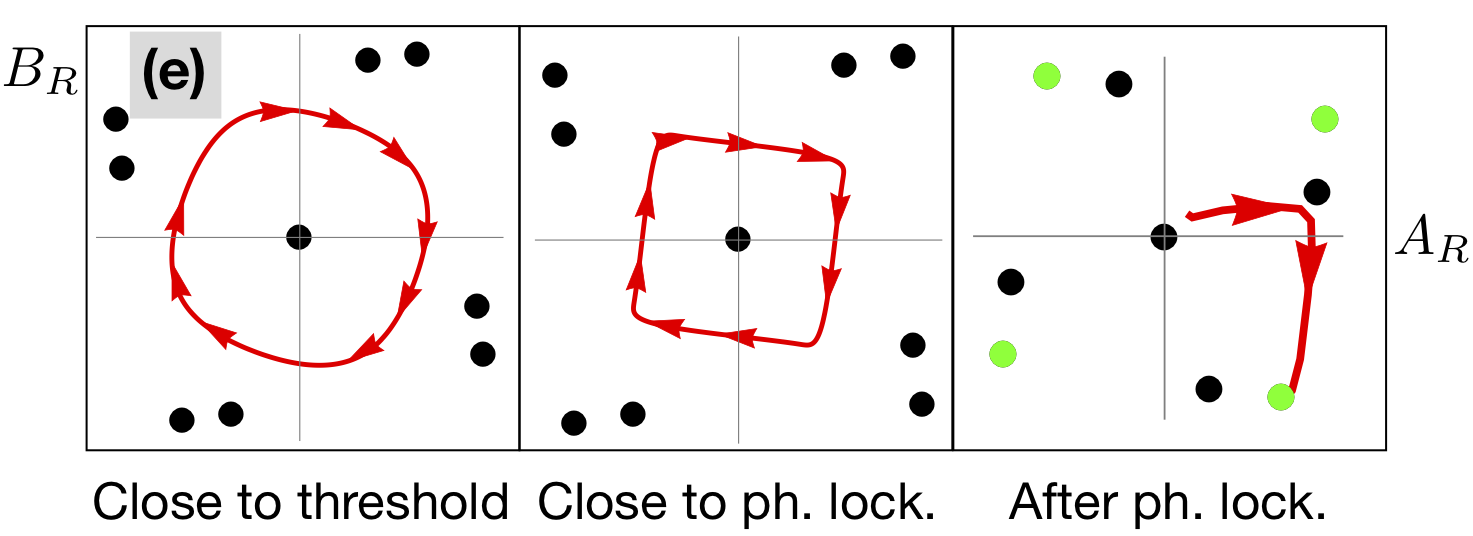}
\caption{(Top panels) Experimental [panels (a), (c)] and numerical [panels (b), (d)] time evolution of the fields $x_1(t)$ (blue) and $x_2(t)$ (red), and corresponding slow-varying amplitudes (orange and black, respectively). The data are taken (a)-(b) just above the oscillation threshold, and (c)-(d) close to \textcolor{black}{phase locking}. (e) Flow of Eq.~\eqref{eq:nonlinearmathieuequationflowcoupled} shown as the real part of $B$ ($B_R$) vs. the real part of $A$ ($A_R$) (red lines). Saddle and stable points are represented by black and green dots, respectively.}
\label{fig:oscillatorstimedependence}
\end{figure}

The fast time scale of the oscillator is associated with the carrier frequency $t\!=\!2\pi/\omega_0$ and the loss $g$ is the small expansion parameter of the theory, allowing to identify the slow time scale $\tau\!=\!gt$. We therefore write $x_{1}(t,\tau)\!=\!A(\tau)e^{i\omega_0t}\!+\!A^*(\tau)e^{-i\omega_0t}$ and $x_{2}(t,\tau)\!=\!B(\tau)e^{i\omega_0t}\!+\!B^*(\tau)e^{-i\omega_0t}$, where $A$ and $B$ are the complex amplitudes of $x_1$ and $x_2$, respectively. \textcolor{black}{By normalizing} $\tilde h\!=\!h/g$, $\tilde r\!=\!r/g$ and $\tilde\alpha\!=\!\alpha/g$, and defining $\tilde\tau=\omega_0\tau$, the long-time dynamics is captured by the set of ODEs~\cite{PhysRevA.100.023835}:
\begin{equation}
\!\!\!\!\!\!\left.\begin{array}{l}
\cfrac{\partial A}{\partial\tilde\tau}=\cfrac{\tilde h}{4}\,A^*-\cfrac{\tilde h\,\beta}{4}\left(3{|A|}^2A^*-A^3\right)-\cfrac{A}{2}+\cfrac{\tilde r{-\tilde\alpha}}{2}\,B=0 \vspace{0.2cm}\\
\cfrac{\partial B}{\partial\tilde\tau}=\cfrac{\tilde h}{4}\,B^*-\cfrac{\tilde h\,\beta}{4}\left(3{|B|}^2B^*-B^3\right)-\cfrac{B}{2}-\cfrac{\tilde r{+\tilde\alpha}}{2}\,A=0
\end{array}\right. \!.
\label{eq:nonlinearmathieuequationflowcoupled}
\end{equation}

We now calculate the phase diagram of Eq.~\eqref{eq:nonlinearmathieuequationflowcoupled} in the $\textcolor{black}{{(h/2g)}^2}$ vs. $r$ plane (see Fig.~\ref{fig:floquetphasediagram}), using tools of nonlinear dynamics~\cite{strogatz2007nonlinear} to determine the number of fixed points and their stability. Below the threshold $h\!<\!h_{\rm th}$, a unique stable fixed point exists at $A\!=\!B\!=\!0$ (the origin). Above the threshold ($h\!>\!h_{\rm th}$) the origin is unstable and two situations are encountered ($\tilde\alpha\!\neq\!0$): For $\tilde r\!<\!\tilde\alpha$, two stable fixed points correspond to \textcolor{black}{two preferred phased-locked configurations - in-phase ($00$ or $\pi\pi$) or anti-phase ($0\pi$ or $\pi0$) depending on the sign of $\tilde\alpha$, in which the oscillators phase-lock with a constant envelope}. {For larger $\tilde h$, two additional stable points correspond to the two additional \textcolor{black}{phase-locked} configurations, as discussed in the analysis of CIMs~\cite{PhysRevA.88.063853}. For $\tilde r \!>\!\tilde\alpha$, one first finds a \emph{stable limit cycle}, which manifests itself as beats in the time evolution of $A$ and $B$. \textcolor{black}{In this region, the relative phase between the two oscillators flips periodically between $0$ and $\pi$}. Only for larger $\tilde h$, the \textcolor{black}{region} with two or four stable fixed points appears. If $\tilde\alpha\!=\!0$ the CIM \textcolor{black}{region} does not exist at all. For $\phi\!>\!0$, the width of the limit cycle region gradually decreases, eventually vanishing at $\phi\!=\!\pi$~\cite{PhysRevA.100.023835}.

\textit{Experimental methods}. Since the dynamics described here is coherent and purely classical, it is suitable to realize the coupled parametric oscillators in a radio-frequency (RF) configuration. Although an RF parametric amplifier at room temperature will not demonstrate quantum squeezing, it can realize easily semiclassical squeezing of the classical thermal noise within the oscillator (to be reported in a future publication). Furthermore, an RF experiment is technically very simple and allows us to observe the oscillation also directly \emph{in time} (on an oscilloscope), which is a great advantage compared to optical realizations. 

The coupled parametric oscillators are realized with two ring RF resonators (see Fig.~\ref{fig:schemeoftheexperimentalapparatus}) of \textcolor{black}{70~cm} long coaxial cables with a repetition rate of roughly 85 MHz. Each resonator includes: \textcolor{black}{(A)} an RF frequency mixer pumped at 170~MHz by an RF synthesizer acting as the nonlinear parametric amplifier, \textcolor{black}{(B)} a broadband (regular) low-noise amplifier with gain of approximately 15~dB, which compensates for the losses of the cavity, \textcolor{black}{(C)} a $-15$~dB coupler for the resonator output, and \textcolor{black}{(D)} a tunable attenuator to electronically tune the overall gain of the oscillator. The coupling between the parametric oscillators is achieved with a fixed power splitter and a couple of tunable attenuators to control the effective coupling. The oscillators are pumped by two phase-locked synthesizers, allowing us to control the relative phase between the pumps \textcolor{black}{(see supplemental material for more details)}.

Since we aim primarily at demonstrating the properties of the beating \textcolor{black}{regime} (limit cycle) with energy-preserving coupling, we mostly focus experimentally on $\alpha\!=\!0$ and monitor the field emitted from the parametric oscillators for various values of the pump power $h$ with respect to the oscillation threshold $h_{\rm th}$ and various coupling strengths $r$, determined by the beat frequency at threshold. {Our results are shown in Fig.~\ref{fig:oscillatorstimedependence}(a)-(d). The left plots show the experimental results, while the right panels show the corresponding theoretical solution, obtained by numerically solving Eq.~\eqref{eq:equationsofmotions}. The latter plots are overlapped by the oscillation envelopes $2|A(gt)|$, $2|B(gt)|$ (orange and black), computed by solving the slow-varying Eq.~\eqref{eq:nonlinearmathieuequationflowcoupled}. For pumping slightly above threshold, both oscillators demonstrate a regular, nearly sinusoidal beating envelope over a carrier signal at half the pump frequency, which matches the cavity resonance at 87~MHz [Fig.~\ref{fig:oscillatorstimedependence}(a),(b)]. As we further increase the pump power, the period of the beats increases and their shape becomes elongated and pear-shaped [Fig.~\ref{fig:oscillatorstimedependence}(c),(d)], until finally diverging at the transition to a \textcolor{black}{phase-locked} steady-state (not shown).

In Fig.~\ref{fig:oscillatorstimedependence}(e), we show the flow of Eq.~\eqref{eq:nonlinearmathieuequationflowcoupled} as $B_R\!\equiv\!{\rm Re}[B]$ vs. $A_R\!\equiv\!{\rm Re}[A]$, for three different cases: slightly above the oscillation threshold, where all fixed points are saddle points and the limit cycle is nearly a perfect circle around the origin, corresponding to perfect beats; just before \textcolor{black}{phase locking}, where the limit cycle becomes sharper and the beats assume an asymmetric shape; after \textcolor{black}{phase locking}, where stable attractors around the origin stabilize the dynamics.

From the observed fields inside the cavities, we can obtain an experimental phase diagram to be compared to the theoretical behaviour discussed before (Fig.~\ref{fig:floquetphasediagram}, left panel). For a given set of values of \textcolor{black}{${(h/2g)}^2$} and $r$, we superimpose the experimental points on the theoretical map, marking red dots when beats are observed, and blue crosses when \textcolor{black}{phase locking} is observed (using $g$ as a fit parameter \textcolor{black}{$g\!=\!3.2\!\times\!10^{-2}$}). Close to \textcolor{black}{phase-locking}, the system is very sensitive to noise, and the observed behaviour alternates between beats and \textcolor{black}{phase locking}, which limits the precise estimation of the experimental transition line.

\begin{figure}[t]
\centering
\includegraphics[width=8.7cm]{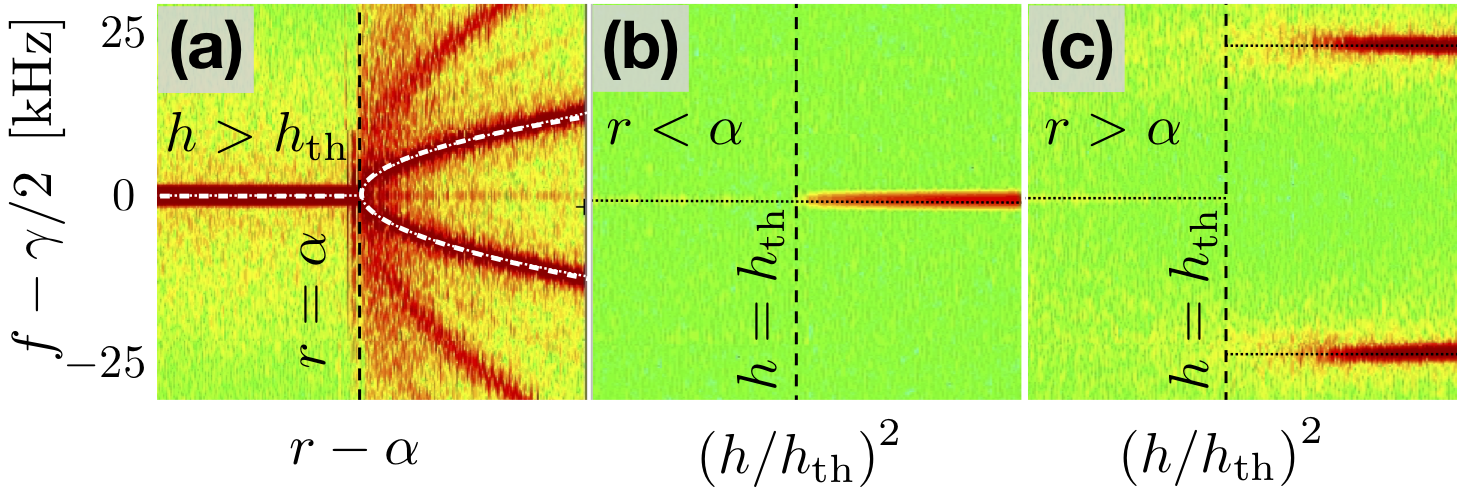}
\caption{\textcolor{black}{Experimental spectrogram of the field inside one of the oscillators. The colormap shows the signal intensity in a log-scale: (a) for a fixed $h\!>\!h_{\rm th}$, the system undergoes a transition from the \textcolor{black}{phase-locked} to the beating \textcolor{black}{regime} at $r=\alpha$. (b) For a fixed $r\!<\!\alpha$, the system enters the \textcolor{black}{phase-locked regime} as the pump crosses the threshold ($h=h_{\rm th}$), whereas (c) for $r\!>\!\alpha$, the system instead enters the beating \textcolor{black}{regime} directly above threshold.} \textcolor{black}{Panels (b) and (c) represent the situations depicted by arrows \textbf{A} and \textbf{B} in Fig.~\ref{fig:floquetphasediagram}, respectively.}}
\label{fig:spectrogramexperiment}
\end{figure}

\textcolor{black}{We now consider experimentally the case of $\alpha\!\neq\!0$. Unfortunately, we cannot produce a quantitative experimental map for both $r\!,\!\alpha\!\neq\!0$ due of imperfections of the mixers in the resonator, which prevent accurate and independent calibration of both $r$ and $\alpha$ when their values are comparable. We can however obtain a spectrogram of the beating fields, monitoring the frequency $f-\gamma/2$ (the offset from the center carrier) as $r\!-\!\alpha$ is scanned from positive to negative. This shows the collapse of the beats exactly at $r\!=\!\alpha$, as shown in Fig.~\ref{fig:spectrogramexperiment}(a), where the pump is fixed slightly above threshold. We observe a \textcolor{black}{phase-locked} state for $r\!<\!\alpha$, whereas for $r\!>\!\alpha$ the spectrum splits into two main symmetric branches (indicating beats). The observed scaling of the main branch (dashed white line) is consistent with the square-root scaling predicted in Eq.~\eqref{eq:nonlinearmathieuequationflowcoupled}~\cite{PhysRevA.100.023835}. The additional branches in the spectrogram for $r\!>\!\alpha$ are due to the anharmonicity of the beats close to the transition (see also Fig.~\ref{fig:oscillatorstimedependence}). Panels (b) and (c) show $f-\gamma/2$ as a function of ${(h/h_{\rm th})}^2$ for $r\!<\!\alpha$ and $r\!>\!\alpha$, respectively. For $r\!<\!\alpha$, the system undergoes a direct transition from below threshold (no signal for $h\!<\!h_{\rm th}$) to \textcolor{black}{phase locking} (at $f\!=\!\gamma/2$), whereas for $r\!>\!\alpha$, crossing threshold leads directly to the beating state with two symmetric frequency components \textcolor{black}{$f-\gamma/2\!=\!\pm\omega_0\sqrt{r^2-\alpha^2}/2$}}.

In conclusion, we reported a detailed study of two coupled parametric oscillators, explored in an RF experiment, analytically and numerically. A single parametric oscillator, which spontaneously breaks the symmetry associated with the time-periodicity of the pump, is the prototype example of a discrete time crystal, analogous to an Ising spin. Although naively, one would expect this to hold also when several parametric oscillators are coupled, our study reveals a much richer phase diagram with a new limit-cycle \textcolor{black}{region}, where the oscillators perform coherent beats that never decay or decohere when the coupling contains a significant energy-preserving component. This beating \textcolor{black}{regime} represents a new class of coherent dynamics that was not previously considered within the vastly researched subject of coupled oscillators and is unique to coupled parametric oscillators, demonstrating a new aspect of their coherent link to the pumping field and to each other. 


\begin{acknowledgements}
We are grateful to J. Avron, I. Bonamassa, C. Conti, N. Davidson, I. Gershenzon, D. A. Kessler, R. Lifshitz, M. E. Meller, Y. Michael, C. Tradonsky and Y. Yamamoto for fruitful discussions. A.~P. acknowledges support from ISF grant No.~44/14 and BSF-NSF grant No. 2017743.  M.~C.~S. acknowledges support from the ISF, grants No.~231/14 and~1452/14.
\end{acknowledgements}



%

\onecolumngrid
\pagebreak
\begin{center}
\textbf{\large Supplemental Material for ``Persistent coherent beating in coupled parametric oscillators''}
\end{center}
\setcounter{equation}{0}
\setcounter{figure}{0}
\setcounter{table}{0}
\setcounter{page}{1}
\makeatletter
\renewcommand{\theequation}{S\arabic{equation}}
\renewcommand{\thefigure}{S\arabic{figure}}
\renewcommand{\bibnumfmt}[1]{[S#1]}
\renewcommand{\citenumfont}[1]{S#1}

\section{Details on the Experimental Apparatus}
\label{sec:details_of_experiment}

The experiment was realized with a radio-frequency (RF) configuration using off-the-shelf components. 
As mentioned in the main text, each parametric oscillator was comprised of (Fig.~\ref{fig:experimental_diagram}):\
\begin{itemize}
\item A coaxial cable acting as the resonator;
\item A frequency mixer ($\otimes$) pumped by an external signal generator ($h$) to provide the parametric gain;
\item A linear amplifier (G) to compensate losses;
\item An output coupler (OC) used to observe the output of each oscillator (OUT);
\item A variable-coupling mechanism (red box) to induce the coupling between the oscillators.
\end{itemize}
In this appendix, we report on each of the parts in detail.

Each resonator was comprised of an approximately $70\,{\rm cm}$-long coaxial cable. The length of the cable dictated the repetition rate and resonant frequencies of the oscillator (a longer cable would correspond the a higher repetition rate and a higher density of resonant modes). For this experiment, a relatively short resonator was convenient, since it allowed us to work with a single isolated mode. A cable length of approximately $70\,{\rm cm}$ corresponded to a repetition rate of approximately $85\,{\rm MHz}$. The exact frequency of the resonant mode was inconsequential, since the pump was tuned to match it.

\begin{figure}[b]
	\centering
	\includegraphics[width=0.58\textwidth]{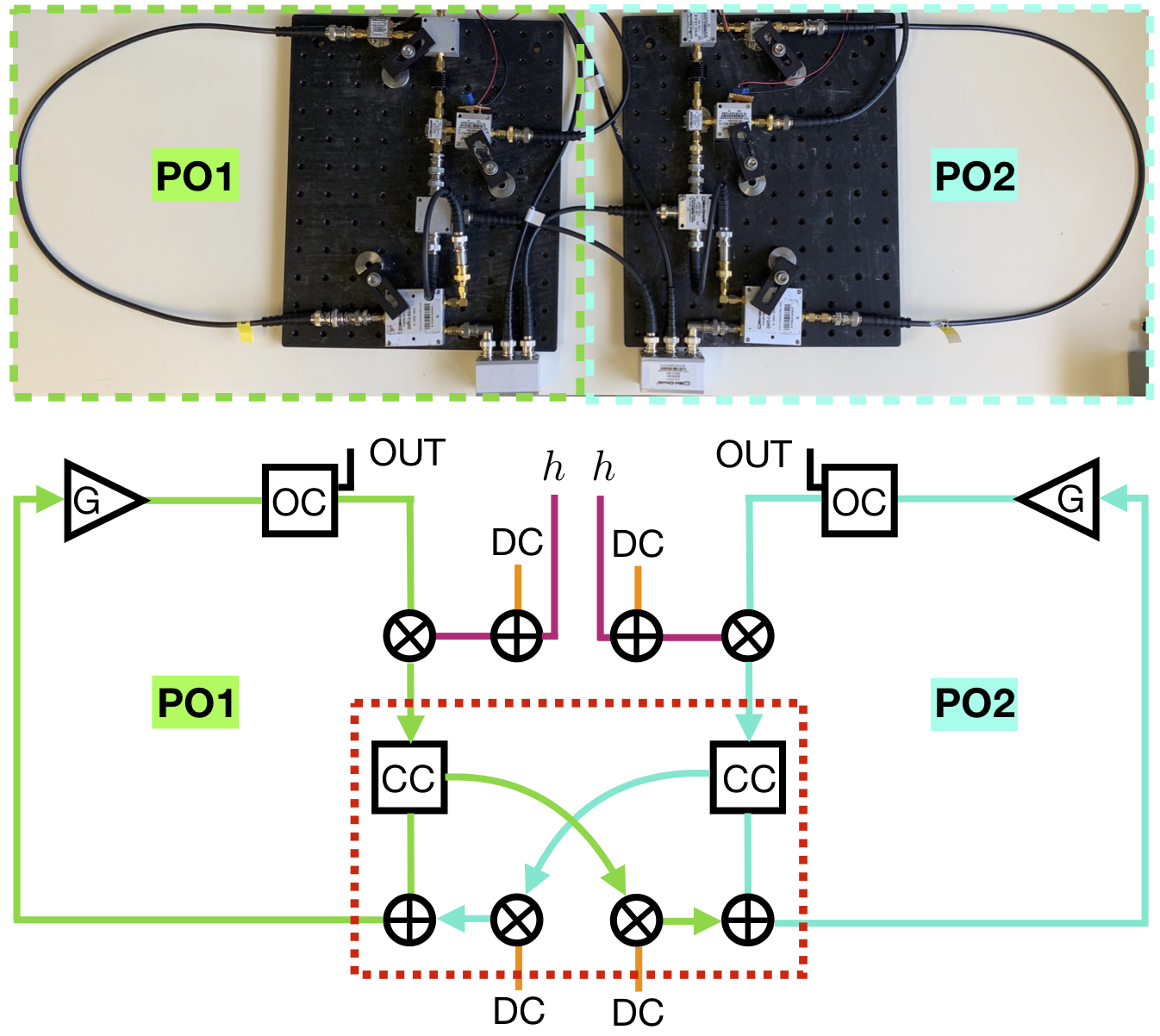}
	\caption{Top figure is a picture of the actual experimental apparatus. Bottom figure is a simplified scheme of the experiment. In the scheme, $\otimes$ stands for a mixer, $\oplus$ stands for a combiner, G denotes a linear amplifier, OC stands for an output coupler and CC for a coupling coupler The orange lines indicate where the DC offset was applied. The purple lines indicate where the pump field with amplitude $h$ was injected. The variable coupling mechanism is boxed with a red dashed line.}
	\label{fig:experimental_diagram}
\end{figure}

The parametric amplifier was realized using a frequency mixer (Mini-Circuits ZX05-10-S+) driven by an external signal generator (Agilent N5181A) at approximately $170\,{\rm MHz}$. Usually, frequency mixers are balanced such that an incident field is (ideally) entirely converted to another frequency, but this was undesirable for our purposes since it would have completely changed the resonances of the oscillators. For our purposes, the mixers needed to be unbalanced, and was used in a non-standard way where the pump is input with a constant DC offset at the intermediate-frequency (IF) port. In addition, this had the added benefit of letting us delicately change the amount of attenuation in the cavity.

The gain of the frequency mixer was generally not enough to cross the oscillation threshold, and thus a broadband (linear) low-noise amplifier (G) (Mini-Circuits ZX60-P105LN) was also added. In order to ensure that oscillation is indeed due to a parametric instability and not due to a lasing one, we worked at an operating point where the parametric gain was substantial and the total linear gain was small and could not induce oscillations on its own.

The output of each oscillator was coupled out using an output coupler (OC) (Mini-Circuits ZFDC-15-5), which we then observed on a scope or a spectrum analyzer, depending on the physical observable that we wanted to measure.

Finally, as mentioned, the two oscillators were coupled with a variable-coupling mechanism. A constant amount of power was coupled out from each oscillator, passed through a DC-controlled variable attenuator and then coupled back into the other oscillator using a power combiner (Mini-Circuits ZAPD-2-252-S+), denoted by $\oplus$ in the figure. By independently changing the amount of attenuation in each of them, we changed the effective value of the coupling parameters $r\pm\alpha$. The variable attenuator was implemented using a DC-offset frequency mixer, which had the added benefit of letting us control the sign of  the coupling. The phase-shift between the pumps could be controlled by detuning them by a very small amount $\delta f$ and then tuning them back to the same frequency after a time $\delta t = \frac{1}{2\pi} \frac{\delta \phi}{\delta f} $.

\end{document}